# Deep-Learning-based Vasculature Extraction for Single-Scan Optical Coherence Tomography Angiography

Jinpeng Liao[1], Tianyu Zhang[1], Yilong Zhang[1], Chunhui Li[1] and Zhihong Huang[1]

[1] School of Science and Engineering, University of Dundee, Dundee, DD1 4HN (UK)

**Abstract.** The skin microvasculature plays a crucial role in physiopathological conditions, constituting a vital aspect of overall skin function and health. Optical coherence tomography angiography (OCTA) is a non-invasive imaging modality for analyzing skin microvasculature, enabling more accurate diagnosis and treatment monitoring. Traditional OCTA algorithms necessitate multi-repeated scans (e.g., 4-8 repetitions) to produce high-quality OCTA images. However, a higher repetition count increases data acquisition time, causing patient discomfort and more unpredictable motion artifacts, which can result in low-quality OCTA images and potential misdiagnosis of skin diseases. In this study, we proposed a vasculature extraction pipeline based on the vasculature extraction transformer (VET) to generate OCTA images using a single OCT scan. Distinct from the Vision Transformer, VET utilizes convolutional projection to better learn the spatial relationships between image patches. Our results show that in comparison to OCTA images obtained via the SV-OCTA (PSNR: 17.809) and ED-OCTA (PSNR: 18.049) using four repeated OCT scans, OCTA images extracted by the proposed pipeline exhibit similar quality (PSNR: 17.515) and higher image contrast while reducing the data acquisition time by 75%. In visual comparison, the proposed pipeline outperforms traditional OCTA algorithms in the neck and face OCTA data that are challenging to scan. This study firstly represents that the VET has the capacity to extract vasculature images from a fast single OCT scan, thereby facilitating accurate diagnosis for patients.

## 1    Introduction

The microvasculature system of skin is considered a representative vascular bed for assessing systemic microvascular reactivity, contains a large amount of information about skin and systemic disease [1]–[3]. For instance, decreased microvascular density has been associated with cardiovascular and metabolic diseases, such as hypertension, diabetes, obesity, and metabolic syndrome, as well as an increased risk of coronary artery disease [4]–[7]. Optical coherence tomography angiography (OCTA) is an extension function based on OCT, providing a microvascular image by extracting the moving red blood cells signals from the surrounding relatively static biological tissue signal [8]–[10]. OCTA had proven that can assist to identify skin disease by assessing the distribution of the vasculature [11]. Particularly, OCTA has emerged as a valuable tool for analyzing skin microvasculature, allowing for more accurate diagnosis and treatment monitoring in skin disease and cancer [12]–[14]. As clinical experience with OCTA continues to grow, it is inspiring innovative advancements in the laboratory, concentrating on improving image quality and expediting acquisition speeds, which will further diversify clinical applications of OCTA.

Among the conventional OCTA algorithms that utilize the differentiative of information (e.g., phase, and complex information) present in OCT signals, speckle variance (SV)-OCTA [15] and eigen-decomposition (ED)-OCTA [16] are highly efficient methods for extracting vasculature images [9]. The quality of OCTA images obtained using there approaches is highly dependent on the number of repeated OCT scans. A greater number of repetitions at the same position can produce higher quality OCTA images [17]. Specifically, in the context of *in vivo* skin OCTA imaging, several factors can significantly degrade the quality of vascular signals, including speckle noise inherent to the OCT system, the motion artifacts caused by bulk tissue motion, and the light waves scattering due to the complex structure of skin tissue. Although increasing the number of repeated scans can improve the quality of skin OCTA images obtained via the SV-OCTA and ED-OCTA algorithms, a higher repeated number necessitates a longer data acquisition time, resulting in more unpredictable motion artifacts. Moreover, distinct from ophthalmology with a fixed objective lens, *in vivo* skin OCTA scan requires a flexible scanning probe to image different positions of sun-exposed skin (e.g., face, hand, and arm) that are easily present with skin cancer [18]. Therefore, the motion artifacts will be increased due to the patient and sample lens under a long data acquisition time of the high-repeated OCTA scan, which leads to a low-quality resultant of OCTA images.

To simultaneously satisfy the image quality and data acquisition speed of OCTA scan, a series of convolution neural network (CNN)-based methods were proposed to improve the quality of OCTA images generated by two- or four-repeated OCT signals [19]–[21]. These approaches achieved competitive results for low-quality OCTA image reconstruction; however, they require at least two-repeated OCT scans for high-quality OCTA image reconstruction. Furthermore, these approaches focus on the mice's brains with an invasively OCTA scan. Rather than solely concentrating on OCTA image reconstruction, the CNN models have to relearn the different features of skin vasculatures in dermatology. In terms of model architecture, the CNN models cannot meet the requirements of the skin OCTA image for high-quality reconstruction in this study. Since the CNN-based methods are difficult to learn the global and long-term information [22], [23], and have a high dependency on the locality convolution operation. Recently, vision transformer (ViT) has gained attention as an alternative to CNNs for image classification tasks due to their scalability, flexibility, and ability to handle long-range dependencies [24]. In Liu et al. work [25], a hierarchical shift window (Swin)-transformer was proposed and achieved state-of-the-art results in image classification. Based on the Swin-transformer, SwinIR [22] was proposed to reconstruct the high-quality nature images from the counterpart degraded images, and Swin-UNet [23] for medical image segmentation, and both of them achieved better competitive results than the CNN models. ViT and Swin-Transformer architecture use a linear projection layer (also referred to as a fully connected layer) to generate query, key, and value sequences for multi-head self-attention. However, this can result in a significant increase in the number of parameters, which can affect the efficiency and practicality of these models. Besides, the other limitation of the linear projection layer is that it does not take into account the spatial relationships between the patches, which can be important for OCTA image reconstruction in this study.

To address the limitations of OCTA imaging in skin applications and improve the efficiency and performance of deep-learning-based models for OCTA image reconstruction, we have proposed a vasculature extraction pipeline based on the proposed Vasculature Extraction Transformer (VET) in this study. Distinct from the conventional OCTA algorithms and CNN approaches mentioned above, which require at least two-repeated OCT scans, the proposed pipeline aims to extract skin microvasculature images from a single *in vivo* skin OCT scan (i.e., structural images). Regarding the proposed VET that harnesses the power of the convolutional projection [25] and Transformer for vasculature feature extraction, different from the linear projection layer, convolutional projection used a convolution operation to obtain the key, value, and query sequences, providing spatial relationships between the image patches.

Consequently, our study has the following contributions: (1) We proposed a single-scan-based OCTA imaging pipeline that efficiently reduces the data acquisition time by 75% while providing a similar OCTA image quality compared to four-repeated OCTA images generated by ED-OCTA and SV-OCTA algorithms. (2) We proposed a novel VET model that uses convolutional projection to help the model learn the spatial relationships between the image patches. (3) To our best knowledge, this is the first competitive study of neural networks in skin OCTA imaging to extract vasculature images based on a single OCT scan. (4) We evaluate the performance of the proposed pipeline with a flexible scanning probe for four different scan positions of skin.

## 2    Vascularature Extraction Methods

### 2.1    Conventional OCTA Algorithms

**Speckle Variance.** Speckle variance (SV) algorithm based on consecutive B-scans is performed to obtain motion-contrast information, which can be formulated as the (1):

$$Flow_{SV}(x,z) = \frac{1}{NR}\sum_{i=1}^{N}|(A_{i+1}(x,z) - A_i(x,z)|  \quad (1)$$

where *NR* is the repeated number of scans at the same location. $A_i(x,z)$ indicates the amplitude signal in $i$-th B-scans at lateral location $x$ and depth position $z$.

**Eigen Decomposition.** Eigen decomposition (ED) algorithm is following the principle of orthogonality. Orthogonality gave the idea that an autocorrelation matrix, containing noise subspace eigenvectors is orthogonal to the signal eigenvectors. By sup-

44

pressing the eigenvectors with a large numerical value that represents the static tissues, the clarity vascular image is extracted, according to [8]. The procedure is in (2), (3):

$$E \wedge E^H = \sum_{i=1}^{N} \lambda_B(i) e_B(i) e_B^H(i) \qquad (2)$$

where $E = [e_B(1), e_B(2), ..., e_B(N)]$ is the $N \times N$ unitary matrix of eigenvectors, $\Lambda = [\lambda_B(1), \lambda_B(2), ..., \lambda_B(N)]$ is the $N \times N$ diagonal matrix of eigenvalues, and $H$ is the Hermitian transpose. The eigenvalues $\Lambda$ are sorted in descending order. By subtracting the first $k^{th}$ eigenvectors which mainly are tissue signals, the extraction of the vessel signals $X_v$ under $K$-repeat scans OCT signal $X$ can be written as (3):

$$X_v = [I - \sum_{i=1}^{K} e_B(i) e_B^H(i)] X \qquad (3)$$

where $I$ is the identity matrix. $e_B(i)$ is the $1 \times N$ unitary matrix of eigenvectors.

### 2.2  Single-Scan Vasculature Extraction Pipeline

A schematic diagram of the single-scan vasculature extraction pipeline and neural network training pipeline is shown in **Fig. 1**. In the training stage, the input of neural networks is generated based on the first repeat of multi-repeated OCT signals. The high-quality vascular signal for neural network loss calculation is extracted by the all-repeated OCT signal with the ED-OCTA algorithm. In the test stage, the trained network utilizes the structural image generated based on the single-scan OCT signal and output the predicted vascular signal. The data preprocessing for neural networks training and validation will be described in the following paragraph.

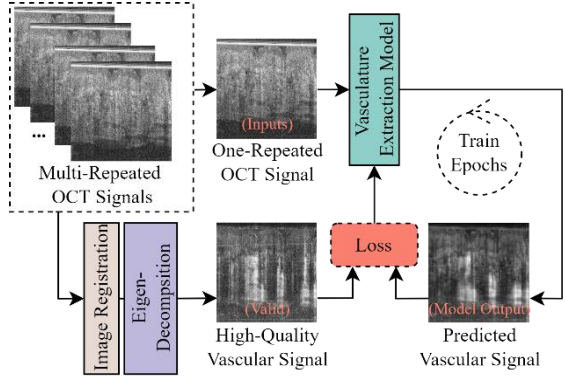

**Fig. 1.** The vasculature extraction pipeline for single-scan OCT image, including the model training pipeline. In the training stage, the predicted vascular signal from the model is used to calculate the loss for the vasculature extraction model's trainable weights updating.

## 2.3 Vasculature Extraction Transformer

Vasculature Extraction Transformer (VET) consists of three modules: shallow feature extraction, residual vasculature feature extraction (RVFE), and feature combination and output block, as shown in **Fig. 2**

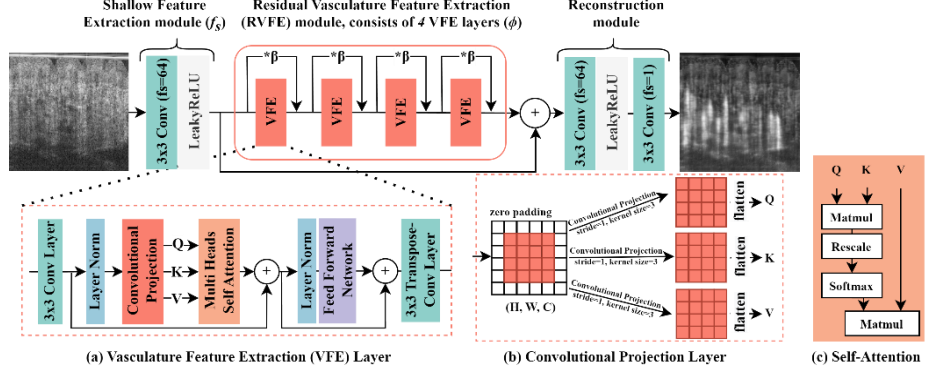

**Fig. 2.** The architecture of the proposed vasculature extraction transformer.

**Shallow feature extraction.** The shallow feature extraction layer ($f_s$) is composed of a 3 × 3 convolution layer (64 filters and strides 1) with a LeakyReLU activation layer. Given a single scan OCT signal (i.e., structural image) input $I_{stru}$ with shape H × W × C, where H, W, and C are image height, width, and channel, respectively, and the processing of the shallow feature extraction layer can be written as:

$$F_s = LeakyReLU(f_s(I_{stru})) \tag{4}$$

where $F_s$ is the obtained shallow feature of the input structural image. According to [26], incorporating an early convolution layer in a transformer architecture model for visual processing can improve optimization stability and lead to improved results.

**Residual vasculature feature extraction.** The residual vasculature feature extraction (RVFE) consists of four VFE layers ($\phi$) and leverages a residual scaling parameter (**β**) to establish an identity connection between VFE layers and the reconstruction module, allowing the aggregation of different levels of features. The forward processing of a VFE layer and a residual connection in RVFE can be written as:

$$F_{out} = F_{in} * \boldsymbol{\beta} + \phi(F_{in}) \tag{5}$$

where $F_{in}$ is the input feature from the previous layer, and $F_{out}$ is the output feature, residual scaling parameter **β** is set as 0.4. The architecture of the VFE layer is illustrated in **Fig. 2** (a), while **Fig. 2** (b) depicts the convolutional projection layer, inspired by [27]. To mitigate the computing cost of multi-head self-attention, in the VFE layer, we employ a 3 × 3 convolution layer ($f_{c1}$) with a stride of 2 that downsamples the input feature ($F_{input}$) shape from H × W × C to H/2 × W/2 × C.

$$F_{c1} = f_{c1}(F_{input}) \tag{6}$$

where $F_{c1}$ is the output downsampled features with shape H/2 × W/2 × C, and $F_{c1}$ is then used as the input of the convolutional projection layer. To ensure both training effectiveness and stability, we opt for a different approach than the squeezed convolutional projection layer used in [27]. Instead, we implement a 3 × 3 convolution projection layer ($f_{CP}$) to obtain query (Q), key (K), and value (V) sequences. This processing procedure (**Fig. 2** (b)) can be formulated as:

$$Q, K, V = Flatten(f_{CP}(LN(F_{c1}))) \tag{7}$$

where $LN$ is the layer normalization layer, and output $Q, K$, and $V$ are then used as the input for multi-head self-attention (MSA). After Flatten processing, the shape of Q, K, and V sequences is (HW/4) × C, and each sequence is split with multi-head by reshaping from (HW/4) × C to M × (HW/4) × C/M, where M is the number of heads. The attention score of each head (M) is then computed using the self-attention mechanism (depicted in **Fig. 2** (c)) as (8). We perform the attention function in parallel M times and concatenate the resulting scores to achieve multi-head self-attention.

$$F_{score} = Attention(Q, K, V) = Softmax\left(\frac{QK^T}{\sqrt{d}}\right) * V \tag{8}$$

where d is a rescale parameter with a numerical value of $1/\sqrt{dims\ of\ Q}$. After the multi-head self-attention operation, the shape of the feature map is (HW/4) × C. Then, a feed-forward network (FFN) that consists of two fully-connected layers with a GELU non-linearity activation layer between them is used for feature transformations. A 2D reshape layer is used to reshape the output of FFN from (HW/4) × C to H/2 × W/2 × C. Finally, a 3 × 3 transpose convolution layer ($f_{tc1}$) with a stride of 2 is used to upscale the shape of the feature map from H/2 × W/2 × C to H × W × C. Generally, the whole process of a VFE layer is formulated as (9) and (10):

$$Y = MSA(f_{CP}(LN(f_{c1}(X)))) + f_{c1}(X) \tag{9}$$

$$Output = f_{tc1}(FFN(LN(Y)) + Y) \tag{10}$$

**Reconstruction Module.** We reconstruct the vascular signal by aggregating the shallow features ($F_s$) from shallow feature extraction module and deep features ($F_{RVEF}$) from residual vasculature feature extraction module:

$$I_V = H_R(F_s + F_{RVEF}) \tag{11}$$

where $I_V$ is the reconstructed vascular signal, and $H_R$ is the reconstruction module as depicted in **Fig. 2**. Shallow features primarily contain low-frequency details, whereas deep features concentrate on recovering lost high-frequency vascular signals. VET utilizes a global skip connection from the shallow feature extraction module to transmit low-frequency information directly to the reconstruction module. This enables the deep feature extraction module to focus on high-frequency information and stabilize training [22].

## 3 Experiment Setup

### 3.1 Data Acquisition and Pre-Processing

A lab-built 200 kHz swept rate swept-source (SS)OCT scan system was utilized to non-invasively collect the OCT data with a hand-held probe, as demonstrated in **Fig. 3**. More details of the SSOCT system were demonstrated in [28]. The data collection of the volunteers was approved by the School of Science and Engineering Research Ethics Committee of University of Dundee, which also conformed to the tenets of the Declaration of Helsinki. All participants had to give their informed consent before entering the lab for the data collection, and the data collected in this article obtained the informed consent of the participants. To develop a comprehensive assessment of the proposed VET, the scan positions were palm and arm (representative 'thick' skin), and face and neck (representative 'thin' skin) taken from fifteen subjects ages between 20 and 35 years old, none of whom had any skin conditions. Moreover, an OCTA scan was also applied to the oral lip position for further investigation.

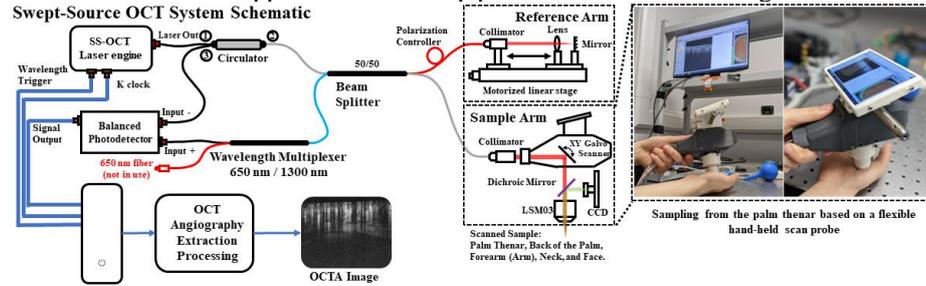

**Fig. 3.** The schematic of the lab-built swept-source optical coherence tomography system. The Laser wavelength is 1310 nm with 100nm bandwidth. The A-scan swept rate is 200 kHz. The flexible hand-held scan probe (sample lens) is demonstrated in the right figure.

In terms of imaging protocol for data acquisition, one OCTA scan can acquire data with a pixel size of NR × 600 × 600 × 300 (NR × x × y × z). Here, NR refers to the number of repeated scans, while x and y represent the transverse axis, and z represents the axial axis. During the OCTA data acquisition, 12 repeated scans were performed for the palm and arm area, 6 for the face and neck areas, and 4 for the lip area. Each repeated scan took approximately 1.8 seconds. The spatial interval in the transverse axis is ~8.6 µm/pixel and ~ 7.4 µm/pixel in the axial axis. After manually removing the low-quality and high-motion artifacts data, we finally collected a total of 25 OCT raw data (11 for palm, 5 for face, 6 for neck, 2 for arm, and 1 for lip). 18 raw data (9 for palm, 4 for face, 4 for neck, 1 for arm) were selected to generate train datasets, and the remaining 7 raw data (including 1 lip data) were used for validation.

The flowchart for dataset pre-processing is shown in **Fig. 4**. To better describe the data pre-processing, we define that one OCT raw data consists of NR volumes, and each volume has a size of 1 × 600 × 600 × 300 (1 × x × y × z), where NR is the number of repeated OCT scans. Firstly, all NR volumes are processed by frame-to-frame registration based on the fast Fourier transform (FFT), and then an FFT-based per A-lines alignment is used to reduce the motion artifacts [29], [30]. The ground-truth high quality OCTA images are generated by using all NR volumes with ED-OCTA algorithms mentioned in (3). Since the ED-OCTA has an outstanding performance in suppressing static tissue while preserving vascular signals [31]. The input skin structural images for neural networks are then generated by using only one OCT volume. The baseline OCTA images are obtained by SV-OCTA and ED-OCTA algorithms with the first four OCT volumes. Since the four-repeated OCTA scans are most frequently used in clinical setups, based on the consideration of imaging acquisition efficiency.

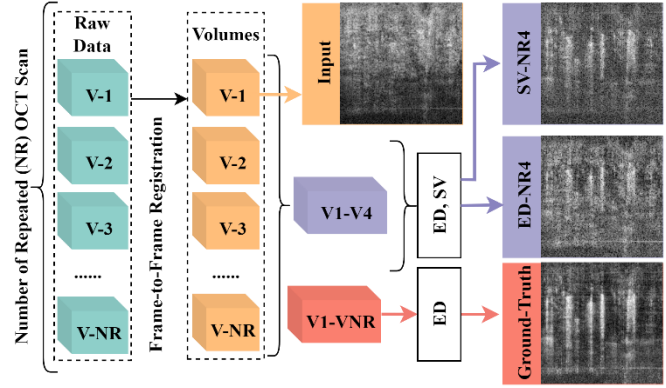

**Fig. 4.** Flow chart of the scanning and processing strategy to create ground-truth OCTA results using twelve-repeated scans, baseline OCTA results using four-repeated scans, and the strategy to obtain input structural images based on a single OCT scan. The frame-to-frame registration consists of fast Fourier transfer (FFT) to obtain structural volume and FFT-based per A-lines alignment (V-1 is used as reference) to reduce the motion artifacts.

After the data pre-processing for all 25 OCT raw data (18 for training and 7 for validation), 15000 B-frames were extracted ($25 \times 600$ frames/data). An image crop box with a size of $192 \times 192$ is then used to extract image patches from each B-frame image. Finally, a total of 45000 pairs of images are generated for the ground-truth, baseline, and input datasets. Among them, 32400 images (from 18 raw data) are used as training datasets for neural network training, and the remaining 12600 images (from 7 raw data) are selected as validation datasets for quantitative comparison.

### 3.2 Implementation Details

The VET is trained based on TensorFlow 2.9.0. To enhance data diversity during the training phase, data augmentation techniques such as flipping and rotations were employed, contributing to the improved generalization of the trained model and mitigating overfitting. The filter size for all convolution layers in the VET is set to 64, with the exception of the final output convolution layer. Within the feed-forward network, the first fully-connected layer comprises 256 hidden units, while the second fully-connected layer contains 64 units. All other aspects of the VET implementation remain consistent with the methodology described in the corresponding section.

The VET model was optimized using an Adam optimizer [32] (with a 0.0001 learning rate, 0.8 for beta1, and 0.999 for beta2) on an Nvidia RTX 3090 GPU with 24GB memory. The training process utilized a batch size of 4 and ran for 200 epochs, using mean-square-error (MSE) as the loss function because it can provide a better performance and training stability over the mean-absolute-error loss function.

### 3.3 Comparison with the Networks

To assess the performance of our proposed VET model for vasculature extraction, we conducted a comparative analysis of the image quality between OCTA images extracted using various neural networks, including DnCNN [33], U-Net [34], SRGAN [35], ESRGAN [36], TransUNet [37], SwinIR [22], and Swin-UNet [23]. The image quality evaluation of the OCTA images was performed both quantitatively and qualitatively. Additionally, we provide the total number of parameters and floating-point operations (evaluated on a 192 × 192 size image).

Notably, SRGAN and ESRGAN were originally designed for natural image super-resolution; therefore, we removed the upsample layers from these two networks. To minimize the influence of network training specifics, we maintained the implementation details for DnCNN, SRGAN, ESRGAN, and SwinIR as per the published sources. As for U-Net, TransUNet, and Swin-UNet, which were initially developed for image segmentation, we utilized the mean squared error (MSE) loss function with supervised training (i.e., the same as the VET implementation details). Regarding the optimizer, epochs, batch size, and data augmentation, all compared networks follow the same configuration as described in Section-3.2.

### 3.4 Evaluation Metrics

To conduct a quantitative performance comparison of various methods, including SV-OCTA, ED-OCTA, and deep-learning-based methods, this study utilized peak-signal-to-noise ratio (PSNR), structural similarity (SSIM) [38] and multi-scale (MS)-SSIM [39] as objective evaluation metrics. Additionally, to offer a more comprehensive analysis of the vasculature extraction performance, we utilized enface OCTA images generated using the maximum intensity projection (MIP) for visual comparison. These enface OCTA images were compared against a baseline image (**Fig. 4** purple blocks) to assess the performance of the methods in terms of vascular connectivity and vasculature extraction. This visual evaluation approach provided an additional perspective to complement the quantitative analysis, allowing for a more nuanced and accurate assessment of the extraction methods.

## 4 Results

After training all of the networks including the proposed VET model and compared-used networks, we then applied them to extract vascular signals from a set of test data mentioned in Section-3.1. The quantitative comparison is based on the test dataset, and the visual comparison is based on the enface OCTA images generated by different methods. In this section, we discuss the advantages of using neural networks for single-scan OCTA image generation.

### 4.1 Quantitative Comparison between the Different Methods

**Table 1** demonstrates a quantitative comparison of different methods, with all methods improving the image quality of single-repeated structural OCT images in terms of PSNR, SSIM, and MS-SSIM performance. The ED-OCTA method with four-repeated scans achieves the best performance in terms of PSNR (18.049), SSIM (0.374), and MS-SSIM (0.730). Regarding the comparison between the neural networks, VET has the third smallest FLOPs (27.57G) and has the best performance in terms of PSNR (17.515) and SSIM (0.298). SwinIR has the best MS-SSIM (0.584) performance and second-best PSNR (17.5), but the SSIM (0.276) result is relatively low and has the second-largest FLOPs (103.5G). TransUNet and SwinUNet have similar performances, with TransUNet showing better PSNR (17.417 > 17.387) and MS-SSIM (0.562 > 0.539) but a worse SSIM (0.295 < 0.287). Among the CNN models, SRGAN achieves the best SSIM (0.266) and MS-SSIM (0.558) performance, and second-highest PSNR (17.147), while the FLOPs is relatively small (41.68 G).

**Table 1.** Quantitative Comparison of the vasculature images (Mean ± Standard Deviation) Extracted by Different Methods

| Methods | #Params | #FLOPs | #Repeat | PSNR | SSIM | MS-SSIM |
|---|---|---|---|---|---|---|
| Inputs | N/A | N/A | 1 | 8.876 ± 1.711 | 0.018 ± 0.014 | 0.039 ± 0.015 |
| SV-OCTA [15] | N/A | N/A | 4 | 17.809 ± 1.003 | 0.367 ± 0.040 | 0.718 ± 0.037 |
| ED-OCTA [16] | N/A | N/A | 4 | 18.049 ± 1.016 | 0.374 ± 0.040 | 0.730 ± 0.036 |
| DnCNN [33] | 0.557 M | 40.92 G | 1 | 17.215 ± 1.370 | 0.248 ± 0.029 | 0.537 ± 0.047 |
| SRGAN [35] | 0.567 M | 41.68 G | 1 | 17.147 ± 1.405 | 0.266 ± 0.028 | 0.558 ± 0.050 |
| ESRGAN [36] | 3.506 M | 258.5 G | 1 | 15.730 ± 1.685 | 0.242 ± 0.030 | 0.525 ± 0.046 |
| U-Net [34] | 34.56 M | 59.88 G | 1 | 16.434 ± 1.646 | 0.260 ± 0.033 | 0.521 ± 0.069 |
| TransUNet [37] | 52.35 M | 23.01 G | 1 | 17.417 ± 1.132 | 0.287 ± 0.033 | 0.562 ± 0.050 |
| SwinIR [22] | 1.739 M | 103.5 G | 1 | 17.500 ± 1.598 | 0.276 ± 0.034 | 0.584 ± 0.068 |
| Swin-UNet [23] | 50.28 M | 16.12 G | 1 | 17.387 ± 1.783 | 0.295 ± 0.053 | 0.539 ± 0.101 |
| VET | 0.929 M | 27.57 G | 1 | 17.515 ± 1.619 | 0.298 ± 0.034 | 0.573 ± 0.068 |

Note: #Params is the network parameters representing network size; #FLOPs is floating point operations to compare the computational cost; #Repeat is the number of repeated OCT scans for vasculature extraction.

### 4.2 Visual Comparison Result

Visual results of vasculature extraction by the different methods in various positions, including palm, face, neck, and lip, are demonstrated in this section. The visual comparison and quantitative comparison between the different methods are based on enface images generated by the maximum intensity projection method.

**Fig. 5** demonstrates the visual results based on the skin palm area. The result generated by ED-OCTA (C) has lower contrast and fewer vasculature details, while the SV-OCTA (D) presents more micro-vasculature details but the vascular connectivity of the relatively large vasculature is worse than the (C). In terms of neural networks performance, results from DnCNN (E), SRGAN (F), and TransUNet (I) contain a large number of artifacts, reducing the image quality in terms of visual and quantitative metrics (i.e., PSNR and SSIM). The results from encoder-decoder architecture networks (i.e., UNet (H), TransUNet (I), and Swin-UNet (J)) present poor vascular connectivity, although the results from them have higher contrast. Among them, the results from SwinIR (K) and VET (L) present a better performance in terms of vascular connectivity and contrast. Moreover, the results from the VET (L) have the best quantitative performance (PSNR: 13.66; SSIM: 0.28).

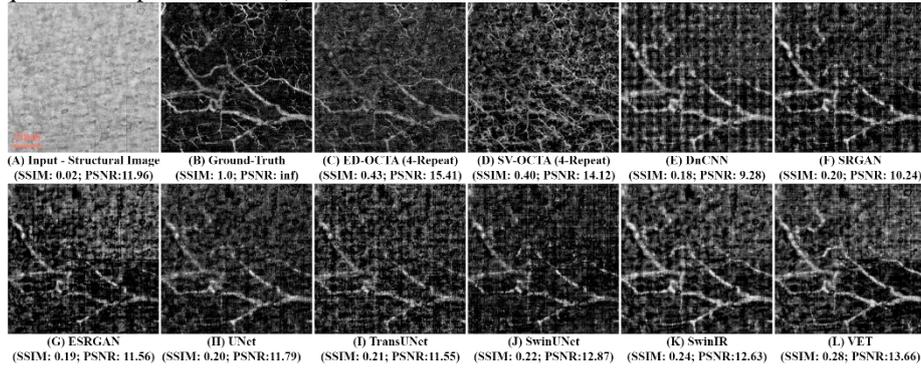

**Fig. 5.** Visual comparison of the hand-held skin palm area. (A) to (L) are enface OCTA images of Input (A), Ground-truth(B), ED-OCTA with four-repeated scan (C), SV-OCTA with four-repeated scan (D), DnCNN (E), SRGAN (F), ESRGAN (G), UNet (H), TransUNet (I), Swin-UNet (J), SwinIR (K), and VET (L). Red scale bar is 1 mm.

**Fig. 6** and **Fig. 7** are visual results based on the neck and face areas, respectively. In **Fig. 6**, the results generated by conventional algorithms (i.e., (C) and (D)) have high motion artifacts and exhibit a lower contrast, compared to the ground truth (B). In the comparison between the neural network results, the U-Net (H), Trans-UNet (I), and Swin-UNet (J) have relatively poor vascular connectivity and vasculature extraction performance. Among them, the results from SRGAN (F), SwinIR (K), and VET (L) have lower motion artifacts than ED-OCTA (C) and SV-OCTA (D), while providing a clearer and relatively higher contrast vasculature extraction result in terms of the visual observation. In **Fig. 7**, the enface OCTA images generated by ED-OCTA (C) and SV-OCTA (D) perform worse in vasculature extraction, and SV-OCTA (D) shows relatively higher motion artifacts. In this stage, the results from neural networks perform a better vasculature extraction than conventional methods (i.e. (C) and (D)) in terms of visual observation. Among them, the VET (L) has the best performance in PSNR (13.64), and ESRGAN (G) has the highest SSIM (0.27).

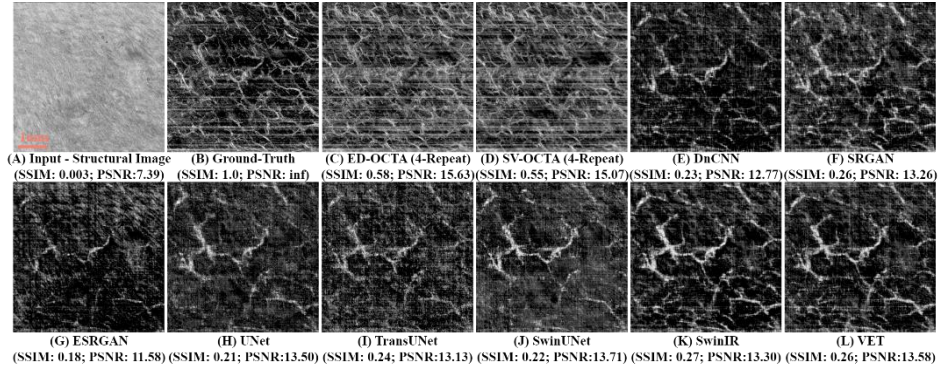

**Fig. 6.** Visual comparison of the hand-held skin neck area. (A) to (L) are enface OCTA images of Input (A), Ground-truth(B), ED-OCTA with four-repeated scan (C), SV-OCTA with four-repeated scan (D), DnCNN (E), SRGAN (F), ESRGAN (G), UNet (H), TransUNet (I), Swin-UNet (J), SwinIR (K), and VET (L). Red scale bar is 1 mm.

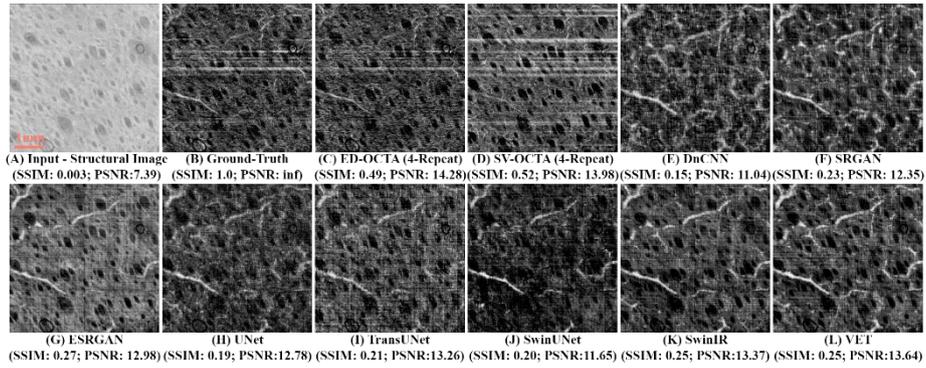

**Fig. 7.** Visual Comparison of the Face Area. (A) to (L) are enface OCTA images of Input (A), Ground-truth(B), ED-OCTA with four-repeated scan (C), SV-OCTA with four-repeated scan (D), DnCNN (E), SRGAN (F), ESRGAN (G), UNet (H), TransUNet (I), Swin-UNet (J), Swin-IR (K), and VET (L). Red scale bar is 1 mm.

The vasculature extraction performance based on the lip OCT scan is represented in **Fig. 8**. It should be noted that the ground truth (i.e., **Fig. 8** (B)) in **Fig. 8** using the four-repeated OCTA scan with ED-OCTA algorithm is due to the lack of six-repeated OCTA scan data. The ED-OCTA (B) result has high contrast and fewer tissue signals. The SV-OCTA (C) result presents more micro-vasculature details but relatively low image contrast. Regarding the neural network's performance, the results from DnCNN (D) and Trans-UNet (H) have high noise and low image contrast. Among them, the SRGAN (E) has the highest SSIM (0.33) but the enface OCTA image consists of the artifacts. While the results from SwinIR (J) and VET (K) perform a higher contrast and less noise. However, the results from all neural networks (i.e., (D)-(K)) perform worse in the microvasculature extraction, compared with the SV-OCTA (C).

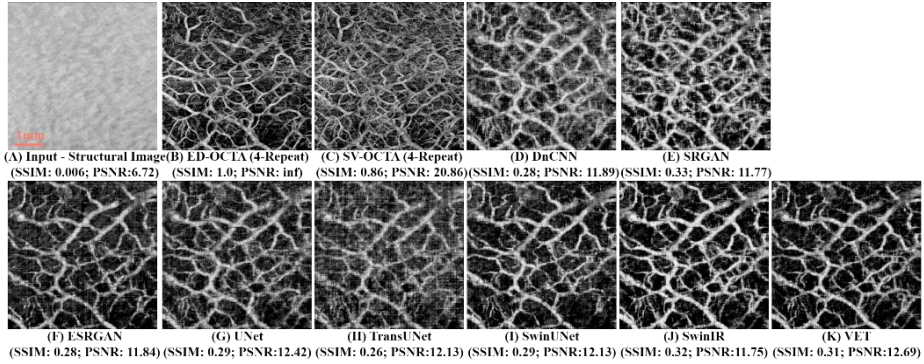

**Fig. 8.** Visual Comparison of the Hand-Held Oral Area. (A) to (K) are enface OCTA images of Input (A), ED-OCTA with four-repeated scan (B), SV-OCTA with four-repeated scan (C), DnCNN (D), SRGAN (E), ESRGAN (F), UNet (G), TransUNet (H), Swin-UNet (I), SwinIR (J), and VET (K). Red scale bar is 1 mm.

## 5 Discussion

In this study, we present a rapid end-to-end vasculature extraction pipeline based on a single *in-vivo* skin OCT scan that requires only ~2 s for data acquisition. Our pipeline employs a novel vasculature extraction transformer (VET), which provides moderate quality OCTA images with a single OCT scan, as opposed to conventional OCTA algorithms like ED-OCTA and SV-OCTA that necessitate at least two-repeated scans. Notably, the VET utilizes convolutional projection to generate query, key, and value sequences for multi-head self-attention computations, preserving spatial relationships between image patches better than fully connected layers used in Trans-UNet, Swin-IR, and Swin-UNet. The results exhibit that our proposed pipeline has significant potential for clinical applications, as it reduces motion artifacts and accelerates imaging speed by reducing the repeated scan of OCTA imaging.

**Table 1** shows a quantitative comparison of different methods. Although neural network results do not surpass ED-OCTA with four-repeated scans, they substantially improve input structural image quality in terms of PSNR (from 8.876 to 17.515), SSIM (from 0.018 to 0.298), and MS-SSIM (from 0.039 to 0.573), as exemplified by the VET. Among deep learning-based approaches, our VET model strikes a balance between the number of parameters (0.929M), FLOPs (26.57G), and performance metrics. Comparing CNN-based and transformer-based models, transformer-based models (i.e., TransUNet, SwinIR, Swin-UNet, VET) generally outperform CNN-based models (i.e., DnCNN, SRGAN, ESRGAN, UNet). Regarding network architecture, end-to-end architectures (e.g., SwinIR, VET) achieve better metrics performance than encoder-decoder architectures (e.g., TransUnet, Swin-UNet). Nevertheless, transformer-type models with encoder-decoder architectures offer smaller FLOPs.

Visual inspection of **Fig. 5** reveals that results from encoder-decoder type networks (i.e., U-Net, TransUNet, SwinUNet) exhibit worse vasculature connectivity and fewer vasculatures details than end-to-end architecture networks (i.e., DnCNN, SRGAN,

SwinIR, VET). Although VET and SwinIR show similar visual performance, VET boasts higher quantitative performance (PSNR: 13.66 > 12.64; SSIM: 0.28 > 0.24).

**Fig. 6** and **Fig. 7** present vasculature extraction results based on challenging high motion artifact OCTA scans (i.e., neck and face area). SwinIR and VET results exhibit fewer motion artifacts, better vasculature connectivity, and higher image contrast than ED-OCTA and SV-OCTA, which use four-repeated OCTA scans, as motion artifacts due to the scanning probe and participants lead to low-quality OCTA images. Furthermore, the results from SwinIR and VET show better vasculature extraction and connectivity than the ground truth, based on visual performance. However, all neural network results struggle with microvasculature extraction when compared to high-quality ground truth images. **Fig. 8** further investigates the vasculature extraction performance of VET in lip OCT data. Despite SRGAN having the highest SSIM (0.33), some vasculatures in its output are not present in ED-OCTA and SV-OCTA. Visually, SwinIR and VET perform better in terms of image contrast and vasculature extraction when compared to ED-OCTA.

Our study has limitations. First, the performance of the proposed VET model may be impacted when using OCT data from diseased subjects, as our data is from healthy participants. In the future, we plan to collect skin OCTA data from participants with various skin conditions and investigate the vasculature extraction pipeline for both healthy and diseased OCT data. Second, we did not apply adversarial training (e.g., generative adversarial network (GAN) [40]) to the VET model training, as it is challenging and can lead to unstable training. We aim to further explore adversarial training for the VET model using conditional GAN [41] and relativistic average (Ra)-GAN [42] to enhance vasculature extraction performance. Third, we acknowledge the importance of conducting an ablation study on the VET model, but due to limited GPU memory, we cannot implement a larger VET model for this study. We plan to address this limitation in our future work by performing a comprehensive ablation study on the VET model to gain a deeper understanding of the contributions of different model components and design choices.

## 6 Conclusion

In this study, we propose an end-to-end vasculature extraction pipeline and VET model that only uses a single OCT scan, demonstrating promising results for clinical applications. The VET model outperforms other deep-learning approaches in terms of efficiency (FLOPs: 27.57G) and performance metrics (PSNR: 17.515; SSIM: 0.298). Despite the limitations in this study, our findings indicate that the proposed pipeline significantly reduces data acquisition time by 75%, while providing similar high-quality OCTA images compared to those obtained by the conventional ED-OCTA algorithm with four-repeated OCT scans. This makes it a valuable tool for fast skin OCTA imaging in clinical settings. In terms of network generalization and robustness, the VET consistently performs stable vasculature extraction across different skin positions (e.g., face, neck, and palm) with varying skin features. In future work, we plan to introduce this fast OCTA scan pipeline to oral and retinal scans, aiming to achieve high-quality OCTA imaging with minimal motion artifacts and rapid acquisition.

**References**

[1]     A. J. Deegan and R. K. Wang, "Microvascular imaging of the skin," *Phys Med Biol*, vol. 64, no. 7, p. 07TR01, 2019.

[2]     M. Roustit and J.-L. Cracowski, "Assessment of endothelial and neurovascular function in human skin microcirculation," *Trends Pharmacol Sci*, vol. 34, no. 7, pp. 373–384, 2013.

[3]     L. A. Holowatz, C. S. Thompson-Torgerson, and W. L. Kenney, "The human cutaneous circulation as a model of generalized microvascular function," *J Appl Physiol*, vol. 105, no. 1, pp. 370–372, 2008.

[4]     M. P. De Boer *et al.*, "Microvascular dysfunction: a potential mechanism in the pathogenesis of obesity-associated insulin resistance and hypertension," *Microcirculation*, vol. 19, no. 1, pp. 5–18, 2012.

[5]     H. Debbabi, L. Uzan, J. J. Mourad, M. Safar, B. I. Levy, and E. Tibiriçà, "Increased skin capillary density in treated essential hypertensive patients," *Am J Hypertens*, vol. 19, no. 5, pp. 477–483, 2006.

[6]     R. G. IJzerman *et al.*, "Individuals at increased coronary heart disease risk are characterized by an impaired microvascular function in skin," *Eur J Clin Invest*, vol. 33, no. 7, pp. 536–542, 2003.

[7]     S. E. Kaiser, A. F. Sanjuliani, V. Estato, M. B. Gomes, and E. Tibiriçá, "Antihypertensive treatment improves microvascular rarefaction and reactivity in low-risk hypertensive individuals," *Microcirculation*, vol. 20, no. 8, pp. 703–716, 2013.

[8]     R. K. Wang, Q. Zhang, Y. Li, and S. Song, "Optical coherence tomography angiography-based capillary velocimetry," *J Biomed Opt*, vol. 22, no. 6, p. 066008, 2017.

[9]     A. Zhang, Q. Zhang, C.-L. Chen, and R. K. Wang, "Methods and algorithms for optical coherence tomography-based angiography: a review and comparison," *J Biomed Opt*, vol. 20, no. 10, p. 100901, 2015.

[10]    A. S. Nam, I. Chico-Calero, and B. J. Vakoc, "Complex differential variance algorithm for optical coherence tomography angiography," *Biomed Opt Express*, vol. 5, no. 11, pp. 3822–3832, 2014, doi: 10.1364/BOE.5.003822.

[11]    B. Zabihian *et al.*, "Comprehensive vascular imaging using optical coherence tomography-based angiography and photoacoustic tomography," *J Biomed Opt*, vol. 21, no. 9, p. 096011, 2016.

[12]    A. J. Deegan *et al.*, "Optical coherence tomography angiography of normal skin and inflammatory dermatologic conditions," *Lasers Surg Med*, vol. 50, no. 3, pp. 183–193, 2018.

[13]    Y. Ji, K. Zhou, S. H. Ibbotson, R. K. Wang, C. Li, and Z. Huang, "A novel automatic 3D stitching algorithm for optical coherence tomography angi-

ography and its application in dermatology," *J Biophotonics*, vol. 14, no. 11, p. e202100152, 2021.

[14] L. Themstrup, G. Pellacani, J. Welzel, J. Holmes, G. B. E. Jemec, and M. Ulrich, "In vivo microvascular imaging of cutaneous actinic keratosis, Bowen's disease and squamous cell carcinoma using dynamic optical coherence tomography," *Journal of the European Academy of Dermatology and Venereology*, vol. 31, no. 10, pp. 1655–1662, 2017.

[15] A. Mariampillai *et al.*, "Speckle variance detection of microvasculature using swept-source optical coherence tomography," *Opt Lett*, vol. 33, no. 13, pp. 1530–1532, 2008.

[16] S. Yousefi, Z. Zhi, and R. K. Wang, "Eigendecomposition-based clutter filtering technique for optical microangiography," *IEEE Trans Biomed Eng*, vol. 58, no. 8, pp. 2316–2323, 2011.

[17] B. Baumann *et al.*, "Signal averaging improves signal-to-noise in OCT images: But which approach works best, and when?," *Biomed. Opt. Express*, vol. 10, no. 11, pp. 5755–5775, Nov. 2019, doi: 10.1364/BOE.10.005755.

[18] M. A. Linares, A. Zakaria, and P. Nizran, "Skin cancer," *Primary care: Clinics in office practice*, vol. 42, no. 4, pp. 645–659, 2015.

[19] X. Liu *et al.*, "A deep learning based pipeline for optical coherence tomography angiography," *J Biophotonics*, vol. 12, no. 10, p. e201900008, 2019.

[20] Z. Jiang *et al.*, "Weakly supervised deep learning-based optical coherence tomography angiography," *IEEE Trans Med Imaging*, vol. 40, no. 2, pp. 688–698, 2020.

[21] Z. Jiang *et al.*, "Comparative study of deep learning models for optical coherence tomography angiography," *Biomed Opt Express*, vol. 11, no. 3, pp. 1580–1597, 2020.

[22] J. Liang, J. Cao, G. Sun, K. Zhang, L. Van Gool, and R. Timofte, "Swinir: Image restoration using swin transformer," in *Proceedings of the IEEE/CVF International Conference on Computer Vision*, 2021, pp. 1833–1844.

[23] H. Cao *et al.*, "Swin-unet: Unet-like pure transformer for medical image segmentation," *arXiv preprint arXiv:2105.05537*, 2021.

[24] A. Dosovitskiy *et al.*, "An image is worth 16x16 words: Transformers for image recognition at scale," *arXiv preprint arXiv:2010.11929*, 2020.

[25] Z. Liu *et al.*, "Swin transformer: Hierarchical vision transformer using shifted windows," in *Proceedings of the IEEE/CVF International Conference on Computer Vision*, 2021, pp. 10012–10022.

[26] T. Xiao, M. Singh, E. Mintun, T. Darrell, P. Dollár, and R. Girshick, "Early convolutions help transformers see better," *Adv Neural Inf Process Syst*, vol. 34, pp. 30392–30400, 2021.

[27] H. Wu *et al.*, "Cvt: Introducing convolutions to vision transformers," in *Proceedings of the IEEE/CVF International Conference on Computer Vision*, 2021, pp. 22–31.

[28] Y. Ji *et al.*, "Deep-learning approach for automated thickness measurement of epithelial tissue and scab using optical coherence tomography," *J Biomed Opt*, vol. 27, no. 1, p. 015002, 2022.

[29]    Y. Cheng, Z. Chu, and R. K. Wang, "Robust three-dimensional registration on optical coherence tomography angiography for speckle reduction and visualization," *Quant Imaging Med Surg*, vol. 11, no. 3, p. 879, 2021.

[30]    S. Klein, M. Staring, K. Murphy, M. A. Viergever, and J. P. W. Pluim, "Elastix: a toolbox for intensity-based medical image registration," *IEEE Trans Med Imaging*, vol. 29, no. 1, pp. 196–205, 2009.

[31]    Q. Zhang, J. Wang, and R. K. Wang, "Highly efficient eigen decomposition based statistical optical microangiography," *Quant Imaging Med Surg*, vol. 6, no. 5, p. 557, 2016.

[32]    D. P. Kingma and J. Ba, "Adam: A method for stochastic optimization," *arXiv preprint arXiv:1412.6980*, 2014.

[33]    K. Zhang, W. Zuo, Y. Chen, D. Meng, and L. Zhang, "Beyond a gaussian denoiser: Residual learning of deep cnn for image denoising," *IEEE transactions on image processing*, vol. 26, no. 7, pp. 3142–3155, 2017.

[34]    O. Ronneberger, P. Fischer, and T. Brox, "U-net: Convolutional networks for biomedical image segmentation," in *International Conference on Medical image computing and computer-assisted intervention*, Springer, 2015, pp. 234–241.

[35]    C. Ledig *et al.*, "Photo-realistic single image super-resolution using a generative adversarial network," in *Proceedings of the IEEE conference on computer vision and pattern recognition*, 2017, pp. 4681–4690.

[36]    X. Wang *et al.*, "Esrgan: Enhanced super-resolution generative adversarial networks," in *Proceedings of the European Conference on Computer Vision (ECCV)*, 2018, p. 0.

[37]    J. Chen *et al.*, "Transunet: Transformers make strong encoders for medical image segmentation," *arXiv preprint arXiv:2102.04306*, 2021.

[38]    Z. Wang, A. C. Bovik, H. R. Sheikh, and E. P. Simoncelli, "Image quality assessment: from error visibility to structural similarity," *IEEE transactions on image processing*, vol. 13, no. 4, pp. 600–612, 2004.

[39]    Z. Wang, E. P. Simoncelli, and A. C. Bovik, "Multiscale structural similarity for image quality assessment," in *The Thrity-Seventh Asilomar Conference on Signals, Systems & Computers, 2003*, Ieee, 2003, pp. 1398–1402.

[40]    I. Goodfellow *et al.*, "Generative adversarial nets," *Adv Neural Inf Process Syst*, vol. 27, 2014.

[41]    M. Mirza and S. Osindero, "Conditional generative adversarial nets," *arXiv preprint arXiv:1411.1784*, 2014.

[42]    A. Jolicoeur-Martineau, "The relativistic discriminator: a key element missing from standard GAN," *arXiv preprint arXiv:1807.00734*, 2018.